\documentclass[aps,pre,reprint,groupedaddress,showpacs]{revtex4-1}
\usepackage[usenames,dvipsnames]{color}
\usepackage{algorithmic}
\usepackage{graphicx}
\usepackage{dcolumn}
\usepackage{cases}

\DeclareGraphicsExtensions{.jpg}

\newcommand{\secref}[1]{section~\ref{#1}}
\newcommand{\Secref}[1]{Section~\ref{#1}}
\newcommand{\appref}[1]{appendix~\ref{#1}}
\newcommand{\figref}[1]{Fig.~\ref{#1}}
\newcommand{\tblref}[1]{Table~\ref{#1}}
\newcommand{\eqref}[1]{Eq.~(\ref{#1})}

\newcommand{\argmax}[0]{\arg\!\max}

\newcommand{\GN}[0]{\citet{GN02}~benchmark }
\newcommand{\LFR}[0]{\citet{LFR08}~benchmark }

\begin{document}


\title{Unfolding communities in large complex networks:\\Combining defensive and offensive label propagation for core extraction}

\author{Lovro \v Subelj}
\thanks{Electronic address: [firstname.lastname]@fri.uni-lj.si}
\author{Marko Bajec}
\thanks{Electronic address: [firstname.lastname]@fri.uni-lj.si}
\affiliation{University of Ljubljana, Faculty of Computer and Information Science, Ljubljana, Slovenia}

\date{\today}

\begin{abstract}
Label propagation has proven to be a fast method for detecting communities in large complex networks. Recent developments have also improved the accuracy of the approach, however, a general algorithm is still an open issue. We present an advanced label propagation algorithm that combines two unique strategies of community formation, namely, defensive preservation and offensive expansion of communities. Two strategies are combined in a hierarchical manner, to recursively extract the core of the network, and to identify whisker communities. The algorithm was evaluated on two classes of benchmark networks with planted partition and on almost $25$ real-world networks ranging from networks with tens of nodes to networks with several tens of millions of edges. It is shown to be comparable to the current state-of-the-art community detection algorithms and superior to all previous label propagation algorithms, with comparable time complexity. In particular, analysis on real-world networks has proven that the algorithm has almost linear complexity, $\mathcal{O}(m^{1.19})$, and scales even better than basic label propagation algorithm ($m$ is the number of edges in the network).
\end{abstract}


\maketitle


\section{\label{sec_intro}Introduction}
Large real-world networks can comprise of local structural modules (\textit{communities}) that are groups of nodes, densely connected within and only loosely connected with the rest of the network. Communities are believed to play important roles in different real-world systems (e.g., may correspond to functional modules in metabolic networks~\cite{PDFV05}); moreover, they also provide a valuable insight into the structure and function of large complex networks~\cite{RB03,PDFV05,LLDM09}. Nevertheless, real-world networks can reveal even more complex modules than communities~\cite{NL07,PSR10}.

Over the last decade the research community has shown a considerable interest in detecting communities in real-world networks. After the seminal paper of \citet{GN02} a vast number of approaches has been presented in the literature. In particular, approaches optimizing \textit{modularity} $Q$ (significance of communities due to a selected \textit{null model}~\cite{NG04})~\cite{CNM04,New06a,BGLL08,BC09,LM09b}, graph partitioning~\cite{RCCLP04,PSSL09} and spectral algorithms~\cite{DM04,New06a}, statistical methods~\cite{NL07}, algorithms based on dynamic processes~\cite{RB08,FCY07,RAK07,LHLC09,RN10}, overlapping, hierarchical and multiresolution methods~\cite{GN02,PDFV05,RN10}, and other~\cite{LM09c} (for an excellent survey see~\cite{For10}). 

The size of large real-world networks has forced the research community in developing scalable approaches that could be applied to networks with several millions of nodes and billions of edges. A promising effort was made by \citet{RAK07}, who employed a simple \textit{label propagation} to find significant communities in large real-world networks. \citet{TK08} have shown that label propagation is in fact equivalent to a large zero-temperature kinetic \textit{Potts model}, when \citet{BC09} have further refined the approach into a modularity optimization algorithm. Just recently, \citet{LM09b} have combined the modularity optimization version of the algorithm with a multistep greedy agglomeration~\cite{SC08}, and derived an extremely accurate community detection algorithm.

\begin{figure}
\includegraphics[width=1.00\columnwidth]{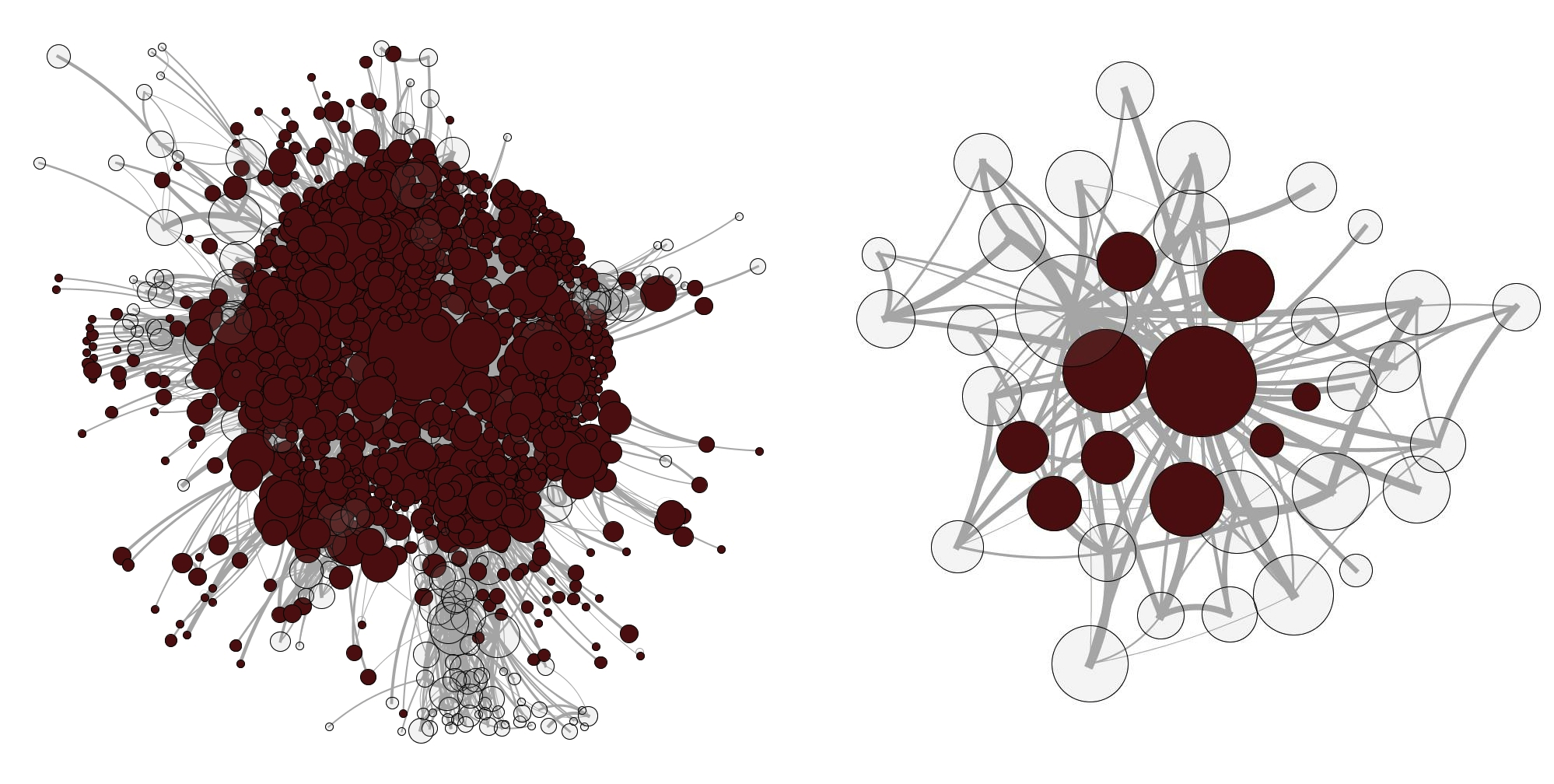}
\caption{\label{fig_asi} (Color online) Results of \textit{diffusion and propagation algorithm} applied to the network of autonomous systems of Internet~\cite{New06b}. Figure shows two \textit{community networks}, where the largest nodes correspond to densely connected modules of almost $10^4$ nodes in the original network. Network cores, extracted by the algorithm, are colored red (dark gray) and whisker communities are represented with transparent nodes. Results show that the algorithm can detect communities on various levels of resolution -- average community sizes are $16.38$ and $588.79$ nodes respectively (with $Q$ equal to $0.475$ and $0.582$ respectively).}
\end{figure}

\citet{LHLC09} have investigated label propagation on large web networks, mainly focusing on scalability issues, and have shown that the performance can be significantly improved with label \textit{hop attenuation} and by applying \textit{node preference} (i.e. node propagation strength). We proceed their work in developing two unique strategies of community formation, namely, \textit{defensive preservation} of communities, where preference is given to the nodes in the core of each community, and \textit{offensive expansion} of communities, where preference is given to the border nodes of each community. Cores and borders are estimated using random walks, formulating the diffusion over the network. 

Furthermore, we propose an advanced label propagation algorithm, \textit{diffusion and propagation algorithm}, that combines the two strategies in a hierarchical manner -- the algorithm first extracts the \textit{core} of the network and identifies \textit{whisker} communities~\footnotemark[1] (\appref{app_cps}), and then recurses on the network core~(\figref{fig_asi}). The performance of the algorithm has been analyzed on two classes of benchmark networks with planted partition and on $23$ real-world networks ranging from networks with tens of nodes to networks with several tens of millions of edges. The algorithm is shown to be comparable to the current state-of-the-art community detection algorithms and superior to all previous label propagation algorithms, with comparable time complexity. In particular, the algorithm exhibits almost linear time complexity (in the number of edges of the network).

The rest of the article is structured as follows. \Secref{sec_lpa} gives a formal introduction to label propagation, and reviews subsequent advances, relevant for this research. \Secref{sec_dpa} presents the diffusion and propagation algorithm and discusses the main rationale behind it. Empirical evaluation with discussion is done in \secref{sec_eval} and conclusion in \secref{sec_conc}.

\footnotetext[1]{Throughout the article we refer to core and whiskers of the network as being the result of the algorithm, although this might not necessarily coincide with the analysis of~\citet{LLDM09}.}


\section{\label{sec_lpa}Label propagation and advances}
Let the network be represented by an undirected graph $G(N,E)$, with $N$ being the set of nodes of the graph and $E$ being the set of edges. Furthermore, let $c_n$ be a community (label) of node $n$, $n\in N$, and $\mathcal{N}(n)$ the set of its neighbors. 

The basic \textit{label propagation algorithm} (LPA)~\cite{RAK07} exploits the following simple procedure. At first, each node is labeled with an unique label, $c_n=l_n$. Then, at each iteration, node is assigned the label shared by most of its neighbors (i.e. \textit{maximal label}), 
\begin{eqnarray}
c_n=\argmax_l|\mathcal{N}^l(n)|,
\label{eq_lpa}
\end{eqnarray}
where $\mathcal{N}^l(n)$ is the set of neighbors of $n$ that share label $l$ (in the case of ties, one maximal label is chosen at random). Due to the existence of multiple edges within the communities, relative to the number of edges between the communities, nodes in a community will adopt the same label after a few iterations. The algorithm converges when none of the labels change anymore (i.e. equilibrium is reached) and nodes sharing the same label are classified into the same community.

The main advantage of label propagation is its near linear time complexity -- the algorithm commonly converges in less then $10$ iterations (on networks of moderate size). \citet{RAK07} observed that after $5$ iterations $95\%$ of nodes already obtain their ``right'' label. Their observation can be further generalized: the number of nodes that change their label on first four iterations roughly follow the sequence $90\%$, $30\%$, $10\%$, $5\%$. However, due to the algorithm's simplicity, the accuracy of identified communities is often not state-of-the-art (\secref{sec_eval}). 

\citet{LHLC09} have noticed that the algorithm, applied to large web networks, often produces a single large community, occupying more than a half of the nodes of the network. Thus, they have proposed a label \textit{hop attenuation} technique, to prevent the label from spreading too far from its origin. Each label $l_n$ has associated an additional score $s_n$ (initially set to $1$) that decreases after each propagation (\eqref{eq_lpa}). Hence,
\begin{eqnarray}
s_n=\left(\max_{i\in\mathcal{N}^{c_n}(n)}s_i\right)-\delta,
\label{eq_delta}
\end{eqnarray}
with $\delta$ being the \textit{attenuation ratio}. When $s_n$ reaches $0$, the label can no longer propagate onward (\eqref{eq_lpaa}), which successfully eliminates the formation of a single major community~\cite{LHLC09}.

\citet{LHLC09} have also shown that hop attenuation has to be coupled with \textit{node preference} $f_n$ (i.e. node propagation strength), in order to achieve superior performance. The label propagation updating rule (\eqref{eq_lpa}) is thus reformulated into
\begin{eqnarray}
c_n=\argmax_l\sum_{i\in\mathcal{N}^l(n)}f_i^\alpha s_iw_{ni},
\label{eq_lpaa}
\end{eqnarray}
where $w_{ni}$ is the edge weight (equal to $1$ for unweighted graphs) and $\alpha$ is a parameter of the algorithm. They have experimented with preference equal to the degree of the node, $f_i=k_i$ and $\alpha=0.1$, however, no general approach was reported.

Label hop attenuation in \eqref{eq_delta} can be rewritten into an equivalent form that allows altering $\delta$ during the course of the algorithm~\cite{LHLC09}. One keeps the label distance from the origin $d_n$ (initially set to $0$) that is updated after each propagation. Hence,
\begin{eqnarray}
d_n=\left(\min_{i\in\mathcal{N}^{c_n}(n)}d_i\right)+1,
\end{eqnarray}
when the score $s_n$ is
\begin{eqnarray}
s_n=1-\delta d_n.
\end{eqnarray}

\citet{RAK07} have already shown that the updating rule of label propagation (\eqref{eq_lpa}), or its refinements (\eqref{eq_lpaa}), might prevent the algorithm from converging. Imagine a \textit{bipartite network} with two sets of nodes, i.e. red and blue nodes. Let, at some iteration of the algorithm, all red nodes share label $l_r$, and all blue nodes share label $l_b$. Due to the bipartite structure of the network, at the next iteration, all red, blue nodes will adopt label $l_b$, $l_r$ respectively. Furthermore, at the next iteration, all nodes will recover their original labels, failing the algorithm to converge.

The problem can be avoided with \textit{asynchronous} updating~\cite{RAK07}. Nodes are no longer updated all together, but sequentially, in random order. Thus, when node's label is updated, (possibly) already updated labels of its neighbors are considered (in contrast to \textit{synchronous} updating that considers only labels from the previous iteration). It should be noted that asynchronous updating can even increase the performance of the algorithm~\cite{LHLC09}.

\begin{figure}
\includegraphics[width=1.00\columnwidth]{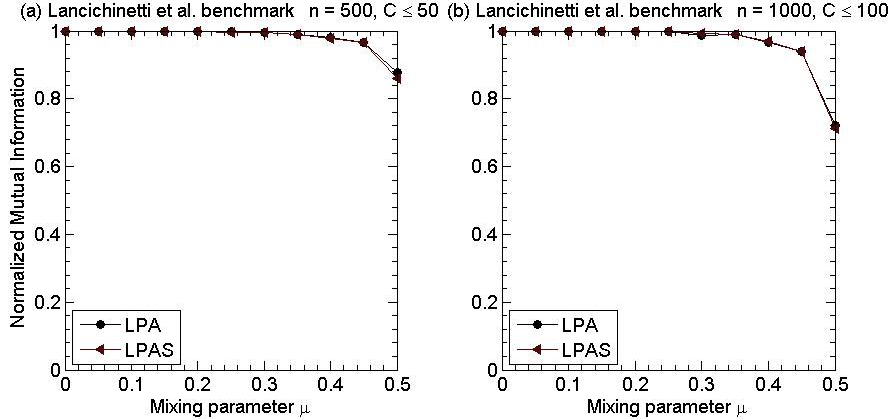}
\caption{\label{fig_eval_random} (Color online) Comparison of node access strategies for label propagation on two sets of benchmark networks with planted partition~\cite{LFR08} (the results are averages over $100$ realizations). Network sizes equal $500$, $1000$ nodes respectively; and communities comprise of up to $50$, $100$ nodes respectively. LPA denotes basic label propagation algorithm and LPAS denotes LPA without (subsequent) reshuffling of nodes.}
\end{figure}

Furthermore, when a node has equally strong connections with two or more communities, its label would, in general, constantly change~\cite{RAK07,LHLC09}. The problem is particularly apparent in author collaboration (co-authorship) networks, where a single author often collaborates with different research communities. On the collaboration network of network scientists~\cite{New06a} the basic label propagation algorithm fails to converge, as there are up to $10\%$ of nodes that would change their label even after $10000$ iterations -- results suggest that there are at least $20\%$ of nodes, i.e. over $300$ scientists, collaborating with different research communities~\footnotemark[2].

\footnotetext[2]{The conclusion naturally depends on the definition of research communities, which are, in this case, communities, revealed by the algorithm.}

\citet{LHLC09} suggested including concerned label itself into the maximal label consideration (and not merely neighbors' labels); however, we use a slightly modified version~\cite{RAK07}. When there are multiple maximal labels among neighbors, and one of them equals the concerned label, the node retains its label. The main difference here is that the modified version considers concerned label only when there exist multiple maximal labels among neighbors. On the discussed collaboration network, such an algorithm converges in around $4$ iterations.

Never converging nodes can also be regarded as a clear signature of \textit{overlapping communities}~\cite{PDFV05}, where nodes can belong to multiple communities. Extension of label propagation to detect overlapping communities was just recently proposed by~\citet{Gre09} (and previously discussed in~\cite{RAK07,LHLC09}). However, due to simplicity, we investigate only basic (no-overlap) versions of the label propagation algorithm.

Another important issue of label propagation is the stability of identified community structure~\cite{RAK07}, especially in large networks. For more detailed discussion see~\cite{RAK07,TK08,LM09b}. 

Label propagation, with asynchronous updating, accesses the nodes in a random order. Nodes are then shuffled after each iteration, mainly to address the problems discussed above. Although this subsequent reshuffling does not increase the algorithm's complexity, it does indeed increase its computational time. Nevertheless, results in \figref{fig_eval_random} show that LPA without subsequent reshuffling of nodes (LPAS) only slightly decreases the performance of the basic LPA. Thus, all the approaches, presented in the following section, use asynchronous updating with a single (initial) shuffling of nodes.


\section{\label{sec_dpa}Diffusion and propagation algorithm}
The section presents \textit{diffusion and propagation algorithm} that combines several approaches, also introduced in this section. We thus give here a brief review of these.

First, we further analyze label hop attenuation for LPA (\secref{sec_lpa}) and propose different \textit{dynamic hop attenuation} strategies in \secref{sec_dpa_att}. Next, we consider various approaches for node propagation preference (\secref{sec_lpa}). By estimating node preference by means of the diffusion over the network, we derive two algorithms that result in two unique strategies of community formation, namely, \textit{defensive preservation} and \textit{offensive expansion} of communities. The algorithms are denoted \textit{defensive} and \textit{offensive diffusion and attenuation} LPA (DDALPA and ODALPA); and are presented in \secref{sec_dpa_dalpa}.

The DALPA algorithms are combined into \textit{basic diffusion and propagation algorithm} (BDPA), preserving the advantages of both defensive and offensive approach (\secref{sec_dpa_main}). BDPA already achieves superior results on networks of moderate size (\secref{sec_eval}), for the use also with larger networks, the algorithm is further enhanced with \textit{core extraction} and \textit{whiskers identification}. The improved algorithm is denoted (general) \textit{diffusion and propagation algorithm} (DPA); and is presented in \secref{sec_dpa_main}.


\subsection{\label{sec_dpa_att}Dynamic hop attenuation}
Hop attenuation has proven to be a reliable technique for preventing the emergence of a major community, occupying most of the nodes of the network~\cite{LHLC09}. It is, however, not evident what should the value of attenuation ratio $\delta$ be (\eqref{eq_delta}). \citet{LHLC09} have experimented with values around $0.10$, and obtained good results, still their experimental setting was rather limited. Furthermore, our preliminary empirical analysis suggests that there is no (simple) universal value for $\delta$, applicable for \textit{all} different types of networks (results are omitted).

\citet{LHLC09} have also observed that large values of $\delta$ may prevent the natural growth of communities and have proposed a dynamic strategy that decreases $\delta$ from $0.50$ towards $0$. In the early iterations of the algorithm, large values of $\delta$ prevent a single label from rapidly occupying large set of nodes and ensure the emergence of a number of strong community cores. The value of $\delta$ is then decreased, to gradually relax the restriction and to allow formation of the actual communities depicted in the network topology. Results on real-world networks show that such a strategy has very good performance on larger networks (\secref{sec_eval}); still, the results can be further improved. Empirical evaluation in \secref{sec_eval} also proves that the strategy is too aggressive for smaller networks, where it is commonly outperformed even by basic LPA.

We propose different \textit{dynamic hop attenuation} strategies, based on the hypothesis that hop attenuation should only be employed, when a community, or a set of communities, is rapidly occupying a large portion of the network. Otherwise, the restriction should be (almost) completely relaxed, to allow label propagation to reach the equilibrium unrestrained. Thus, the approach would retain the dynamics of label propagation and still prevent the emergence of a major community.

We have considered several strategies for detecting the emergence of a large community, or a set of large communities. Due to limited space, we limit the discussion to two. After each iteration, the value of $\delta$ (initially set to 0) is updated according to the following rule:
\begin{description}
\item[\textit{nodes}] $\delta$ is set to the proportion of nodes that changed their label,
\item[\textit{communities}] $\delta$ is set to the proportion of communities (i.e. labels) that disappeared.
\end{description}
Both strategies successfully address the problem of a major community formation, however, the detailed comparison is omitted. The algorithms proposed here all use \textit{nodes} strategy, due to much finer granularity, opposed to the \textit{communities} approach -- after $4$ iterations the number of communities is, in general, already $20$ times smaller than the number of nodes (\secref{sec_lpa}); thus, the estimate of $\delta$ is rather rough for the \textit{communities} strategy. For the empirical evaluation see \secref{sec_eval}.


\subsection{\label{sec_dpa_dalpa}Defensive and offensive propagation}
\citet{LHLC09} have proved that using \textit{node preference}, to increase the propagation strength (i.e. label spread) from certain nodes, can improve the performance of basic LPA. We conducted several experiments by using variations of different measures of node centrality for node propagation preference (i.e. \textit{degree} and \textit{eigenvector centrality}~\cite{Fre77,Fre79} and node \textit{clustering coefficient}~\cite{WS98}). Results are omitted, however, they clearly indicate that none of these static measures applies for \textit{all} different types of networks (i.e. general networks). 

We have also observed that good performance can be obtained by putting higher preference to the core of each community (i.e. to its most central nodes). For instance, on the Zachary's karate club network~\cite{Zac77}, where three high degree nodes reside in the core of the two (natural) communities, degree and eigenvector centralities are superior. However, on \GN networks, where all the nodes have equal degree (on average), the measures render useless and are outperformed by node clustering coefficient. On the \LFR networks, the best performance is, interestingly, obtained by inverted degree or eigenvector centrality. The measures seem to counterpart each node's degree (low degree nodes have high propagation strength, and vice-versa), thus, the propagation utilizes merely the connectedness among nodes, disregarding its strength.

Based (also) on the above observations, we have developed two algorithms that estimate node preference by means of the diffusion over the network. During the course of algorithms, the diffusion is formulated using a \textit{random walker} within each of the (current) communities of the network. The rationale here is twofold: (1) to estimate the (label) propagation within each of the (current) communities~\footnotemark[3]; and (2) to derive an estimation of the core and border of each (current) community (with the core being the most central nodes of the community and the border being its edge nodes).

\footnotetext[3]{The estimation could also be done by using the label propagation itself, however, we believe that using the \textit{floating-point} counterpart of the approach would produce more accurate results in practice.}

Let $p_n$ be the probability that a random walker, utilized on the community labeled with $c_n$, visits node $n$. $p_n$ can be computed as
\begin{eqnarray}
p_n=\sum_{i\in\mathcal{N}^{c_n}(n)}p_i/k_i^{c_n},
\label{eq_diff}
\end{eqnarray}
where the sum goes over all the neighbors of $n$, within the community $c_n$, and $k_i^{c_n}$ is the intra-community degree of node $i$. The employed formulation is similar to the algorithms like \textit{PageRank}~\cite{BP98} and HITS~\cite{Kle99}, and also to the basic eigenvector centrality measure.

Finally, we present the two algorithms mentioned above, namely, \textit{defensive} and \textit{offensive diffusion and attenuation} LPA (DDALPA and ODALPA). The defensive algorithm applies preference (i.e. propagation strength) to the core of each community, i.e. $f_n^\alpha=p_n$, and the updating rule in \eqref{eq_lpaa} rewrites to
\begin{eqnarray}
c_n=\argmax_l\sum_{i\in\mathcal{N}^l(n)}p_is_iw_{ni}.
\label{eq_ddalpa}
\end{eqnarray}
On the other hand, the offensive version applies preference to the border of each community, i.e. $f_n^\alpha=1-p_n$, and the updating rule becomes
\begin{eqnarray}
c_n=\argmax_l\sum_{i\in\mathcal{N}^l(n)}(1-p_i)s_iw_{ni}.
\label{eq_odalpa}
\end{eqnarray}

Opposed to the algorithm of \citet{LHLC09}, the main novelty here is in considering (current) communities, found by the algorithm, to estimate the (current) state of the label propagation process and then to adequately alter the dynamics of the process. 

\begin{figure}
\includegraphics[width=1.00\columnwidth]{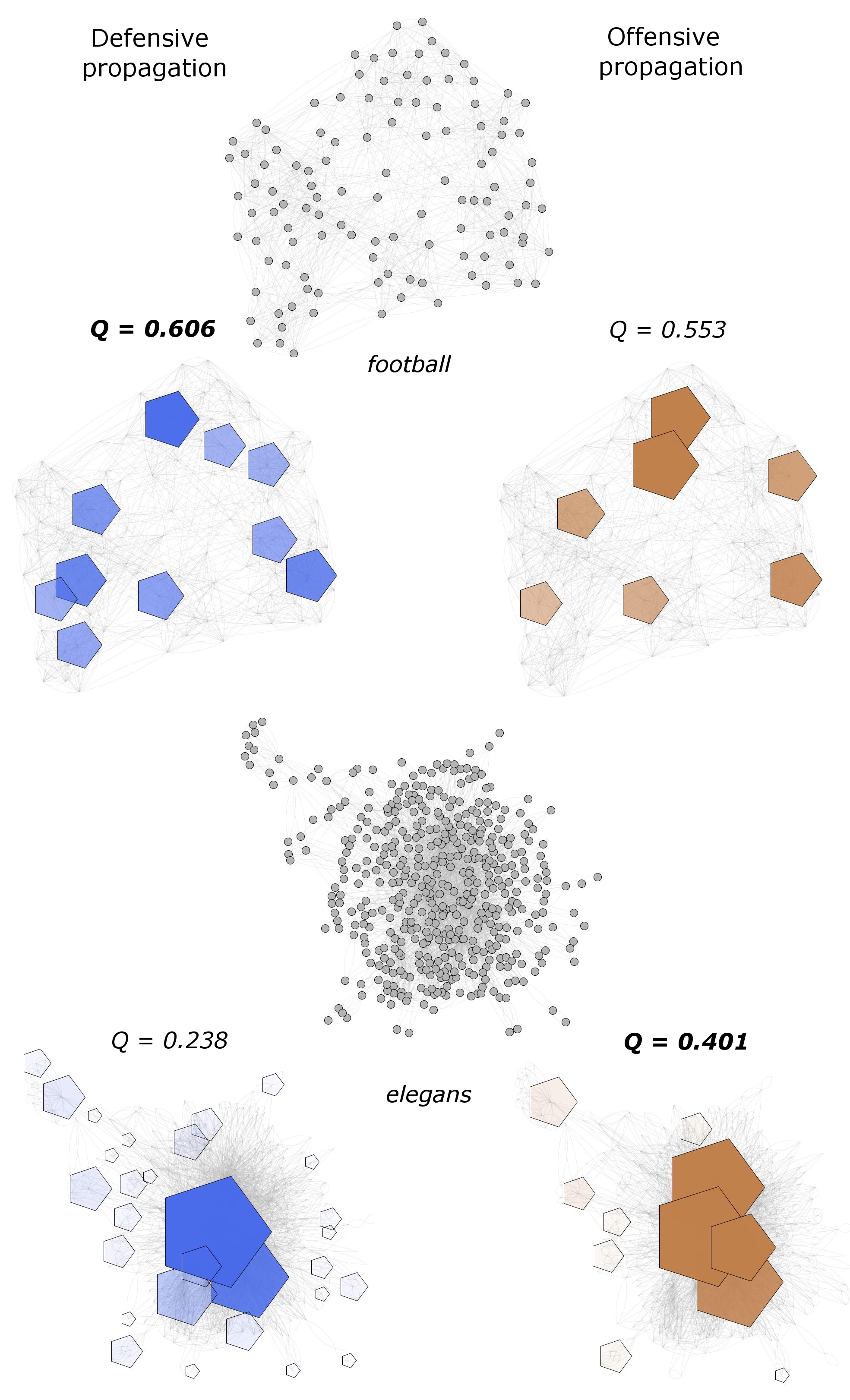}
\caption{\label{fig_dalpa} (Color online) Comparison of defensive and offensive label propagation on two real-world networks, i.e., social network of American football matches on an U.S. college~\cite{GN02} and metabolic network of nematode \textit{Caenorhabditis elegans}~\cite{JTAOB00}. The revealed communities are shown with pentagonal nodes and the sizes, and intensities of colors (shadings), of nodes are proportional to the sizes of communities. The networks comprise two relatively different community structures, considering the distribution of sizes of the communities. That is rather homogeneous in the case of \textit{football} and (presumably) \textit{power-law} in the case of \textit{elegans}.}
\end{figure}

To better estimate the border of each community, the offensive algorithm uses degrees $k_i$ (instead of intra-community degrees $k_i^{c_n}$) for the estimation of diffusion values $p_n$ (see~\eqref{eq_diff}). The modification results in higher values of $1-p_n$ for nodes with large inter-community degrees (i.e. nodes that reside in the borders of communities) and thus provides more adequate formulation of the node propagation strength for the offensive version (result are omitted).

When a node's label changes, the values $p_n$ should be re-estimated for each node in the concerned node's previous or current community. However, this would likely render the algorithm inapplicable on larger networks. Thus, we only update the value $p_n$ (according to \eqref{eq_diff}), when the node $n$ changes its label (initially all $p_n$ are set to $1/|N|$). Although the approach is only a rough approximation of an exact version, preliminary empirical experiments reveal no significant gain by using the exact values for $p_n$.

Defensive and offensive label propagation algorithms result in two unique strategies of community formation, namely, \textit{defensive preservation} and \textit{offensive expansion} of communities. The defensive algorithm quickly establishes a larger number of strong community cores (in the sense of \eqref{eq_ddalpa}) and is able to defensibly preserve them during the course of the algorithm. This results in an immense ability of detecting communities, even when they are only weakly defined in the network topology. On the other hand, the offensive approach produces a range of communities of various sizes, as commonly observed in the real-world networks~\cite{LLDM09,RAK07}. Laying the pressure on the border of each community expands those that are strongly defined in the network topology. This constitutes a more natural (offensive) struggle among the communities and results in a great accuracy of the communities revealed.

Comparison of the algorithms on two real-world networks is depicted in \figref{fig_dalpa}. The examples show that defensive propagation prefers networks with rather homogeneous distribution of the sizes of the communities; and that offensive propagation favors networks with more heterogeneous (e.g \textit{power-law}) distribution. It should, however, be noted that both approaches can achieve superior performance on both of the networks. Still, on average, the defensive approach performs better on social network \textit{football}~\cite{GN02}, when offensive outperforms defensive on the metabolic network \textit{elegans}~\cite{JTAOB00}.

For an empirical analysis and further discussion of the algorithms see \secref{sec_eval}; and for pseudo-code of the algorithms and discussion on some of the implementation issues see \appref{app_algs}.


\begin{figure*}
\includegraphics[width=2.00\columnwidth]{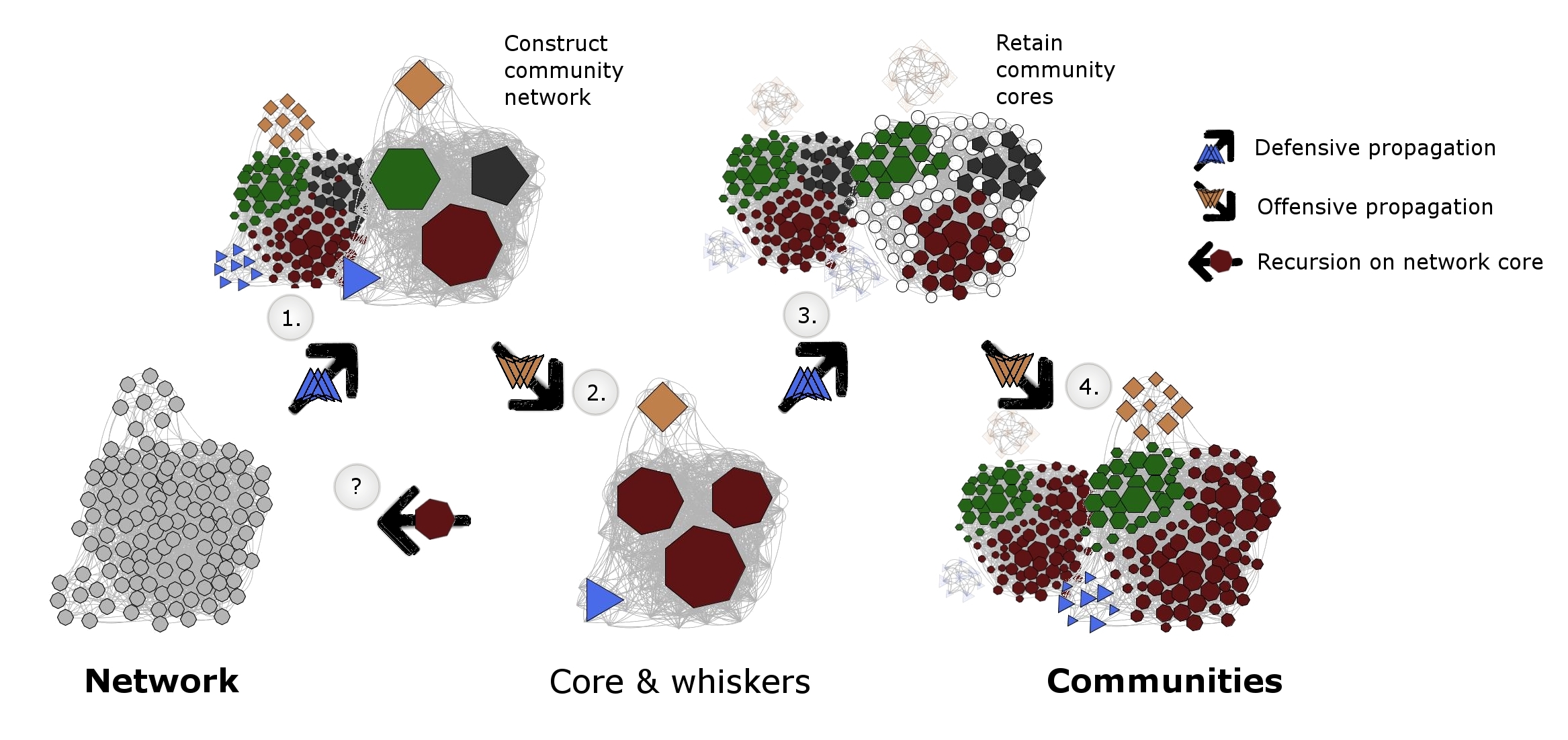}
\caption{\label{fig_dpa} (Color online) Diagram\footnote{Figure is merely a schematic representation of the algorithm and does not correspond to the actual result on the given network.} of (general) \textit{diffusion and propagation algorithm} (DPA). Algorithm combines defensive and offensive label propagation in a hierarchical manner (steps~1.~and~2.), to extract the \textit{core} of the network (red heptagon communities) and to identify \textit{whisker} communities (blue triangle and orange square communities). Whiskers are retained as identified communities, when the algorithm is recursively applied to the core of the (community) network. The recursion continues until all of the nodes of the (current) network are classified into the same community (i.e. offensive propagation in~step~2. flood-fills), when \textit{basic diffusion and propagation algorithm} (BDPA) is applied (steps~3.~and~4.). For more detailed discussion on the algorithms see text.}
\end{figure*}

\subsection{\label{sec_dpa_main}Diffusion and propagation algorithm}
Defensive and offensive label propagation (\secref{sec_dpa_dalpa}) convey two unique strategies of community formation. An obvious improvement would be to combine the strategies, thus, retaining the strong detection ability of the defensive approach and high accuracy of the offensive strategy. However, simply using the algorithms one after another does not attain the desired properties. The reason is that any label propagation algorithm, being run until convergence, finds local optimum (i.e. local equilibrium) that is hard to escape from. 

\citet{RAK07} have already discussed the idea (however, in different context) that label propagation could be improved, if one had \textit{a priori} knowledge about community cores. Core nodes could then be labeled with the same label, leaving all the other nodes labeled with an unique label. During the course of the algorithm, the (uniquely labeled) nodes would tend to adopt the label of their nearest attractor (i.e. community core) and thus join its community. This would improve the algorithm's stability~\cite{RAK07} and also the accuracy of the identified communities (\secref{sec_eval}).

The defensive and offensive label propagation algorithms are combined in the following manner. First, the defensive strategy is applied, to produce initial estimates of the communities and to accurately detect their cores. All border nodes of each community are then relabeled (labeled with an unique label), so that approximately one half of the nodes retain their original label. Last, the offensive strategy is applied, which refines the community cores and accurately detects also their borders. Such combined strategy preserves advantages of both, defensive and offensive, label propagation algorithms and is denoted \textit{basic diffusion and propagation algorithm} (BDPA). Schematic representation of the algorithm is depicted in \figref{fig_dpa} (steps 3. and 4.).

The core (and border) of each community is estimated by means of diffusion $p_n$ (\secref{sec_dpa_dalpa}). As core nodes possess more intra-community edges then border nodes, this results in higher values of $p_n$ for core nodes. Thus, within the algorithm, the node $n$ is relabeled due to the following rule,
\begin{subnumcases}{c_n=}
c_n & for $p_n>m_{c_n}$\\
l_n & for $p_n\leq m_{c_n}$,
\end{subnumcases}
where $m_{c_n}$ is the \textit{median} of values $p_n$, for nodes in community $c_n$, and $l_n$ is an unique label. Thus, the core nodes retain their original labels, when all border nodes are relabeled. Note that all nodes, with $p_n$ equal to median, are also relabeled, to adequately treat smaller communities, where most of the nodes share the same value of $p_n$.

Empirical evaluation shows that BDPA significantly outperforms basic LPA and also the algorithm of~\citet{LHLC09} on smaller networks. However, when networks become larger, the hop attenuation strategy of~\citet{LHLC09} produces much larger communities, with higher values of modularity (on average).

Different authors have proposed approaches that detect communities in a \textit{hierarchical} manner (e.g.~\cite{BGLL08}). The algorithm is first applied to the original network and initial communities are obtained. One then constructs the \textit{community network}, where nodes represents communities, and edges are added between them, when their nodes are connected in the original network. The algorithm is then recursively applied to the community network and the process repeats. At the end, best communities found by the algorithm, are reported (due to some measure).

The idea was also proposed in the context of label propagation~\cite{LHLC09}; however, the authors did not report any empirical results. We have analyzed the behavior of hierarchical label propagation on real-world networks and also on benchmark networks with planted partition. The analysis has shown that, on the second iteration (when the algorithm is first run on the community network), the label propagation (already) produces one major community or even \textit{flood-fills} (all nodes are classified into the same community).

Although the analysis revealed undesirable behavior, we have observed that the major community commonly coincides with the \textit{core} of the network, where other communities correspond to \textit{whisker} communities. \citet{LLDM09} have extensively analyzed large social and information networks and observed that (these) networks reveal clear \textit{core-periphery structure} -- most of the nodes are in the central core of the network that does not have a clear community structure, whereas the best communities reside in the periphery (i.e. whiskers) that is only weakly connected with the core. For further discussion see~\appref{app_cps}.

Based on the above observations, we propose the following algorithm denoted (general) \textit{diffusion and propagation algorithm} (DPA) -- schematic representation of the algorithm is depicted in \figref{fig_dpa}. First, defensive label propagation is applied to the original network (step~1.), which produces a larger number of smaller communities that are used to construct corresponding community network. Second, the offensive label propagation is used on the constructed community network (step~2.), to extract the core of the network (i.e. its major community) and to identify whisker communities (i.e. all other communities). The above procedure is then recursively applied only to the core of the (community) network, when the whisker communities are retained as identified communities. The recursion continues until the offensive propagation in step~2. flood-fills (i.e. extracted core contains all of the nodes of the network analyzed), when the basic BDPA is applied (steps~3.~and~4.).

Empirical analysis on real-world networks shows that DPA outperforms all other label propagation algorithms (with comparable time complexity) and is comparable to current state-of-the-art community detection algorithms. Furthermore, the algorithm exhibits almost linear complexity (in the number of edges of the network) and scales even better than the basic LPA. It should also be noted that the application of the algorithm is not limited to networks that exhibit core-periphery structure.

For a thorough empirical analysis and further discussion on both presented algorithms see \secref{sec_eval}; and for pseudo-code of the algorithms and discussion on some of the implementation issues see \appref{app_algs}.


\begin{figure*}
\includegraphics[width=2.00\columnwidth]{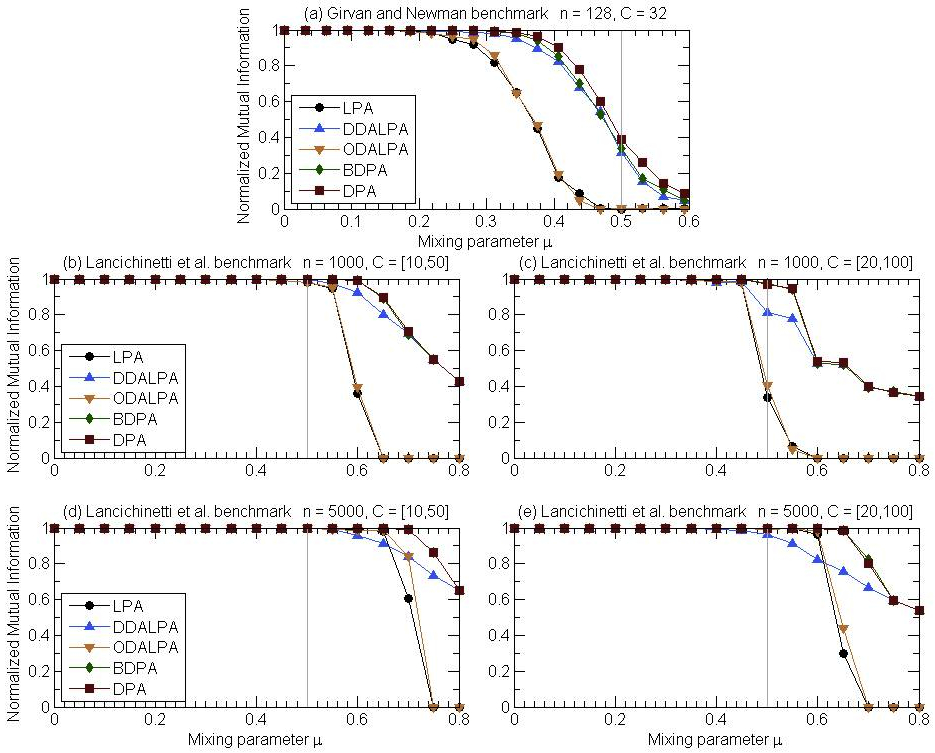}
\caption{\label{fig_eval_benchs} (Color online) Comparison of the proposed algorithms on two classes of benchmark networks with planted partition, namely, \GN networks and four sets of \LFR networks (the results are averages over $100$ realizations). Network sizes equal $128$, $1000$ and $5000$ nodes; and communities comprise of up to $100$ nodes. (Gray) straight lines at $\mu=0.5$ denote the point beyond which the communities are no longer defined in the \textit{strong} sense~\cite{RCCLP04}.}
\end{figure*}

\begin{table*}
\begin{ruledtabular}
\begin{tabular}{cccccccccccccc}
Network & Description & Nodes & Edges & GMO & LPA & LPAD & LPAQ & LPAM & BDPA & DPA & \textit{\# c.e.}\footnotemark[3] & \textit{time}\footnotemark[3] \\\hline
\textit{karate} & Zachary's karate club.~\cite{Zac77} & 34 & 78 & 0.381 & 0.416 & 0.402 & 0.399 & \textbf{0.420} & \textbf{0.419} & \textbf{0.420} & 0.02 \\
\textit{dolphins} & Lusseau's bottlenose dolphins.~\cite{LSBHSD03} & 62 & 159 & & \textbf{0.529} & 0.526 & 0.516 & \textbf{0.529} & \textbf{0.528} & \textbf{0.529} & 0.59 \\
\textit{books} & Co-purchased political books.~\cite{Kre08} & 105 & 441 & & \textbf{0.526} & 0.519 & 0.522 & \textbf{0.527} & \textbf{0.527} & \textbf{0.527} & 0.46 \\
\textit{football} & American football league.~\cite{GN02} & 115 & 616 & 0.556 & \textbf{0.606} & \textbf{0.606} & 0.604 & \textbf{0.605} & \textbf{0.606} & \textbf{0.606} & 0.37 \\
\textit{elegans} & Metabolic network \textit{C. elegans}.~\cite{JTAOB00} & 453 & 2025 & 0.412 & 0.421 & 0.413 & 0.409 & \textbf{\textit{0.452}} & 0.424 & \textbf{0.427}\footnotemark[2] & 0.17 \\
\textit{jazz} & Jazz musicians.~\cite{GD03} & 198 & 2742 & 0.439 & 0.443 & 0.443 & \textbf{0.445} & \textbf{0.445} & \textbf{0.444} & \textbf{0.444} & 0.00 \\
\textit{netsci} & Network scientists.~\cite{New06a} & 1589 & 2742 & & 0.902 & 0.947 & &  & 0.907 & \textbf{0.960} & 1.00 \\
\textit{yeast} & Yeast protein interactions.~\cite{JMBO01} & 2114 & 4480 & & 0.694 & 0.799 &  &  & 0.725 & \textbf{0.824} & 1.04 \\
\textit{emails} & Emails within an university.~\cite{GDDGA03} & 1133 & 5451 & 0.503 & 0.557 & 0.560 & 0.537 & \textbf{\textit{0.582}} & 0.555 & \textbf{0.562} & 0.01 \\
\textit{power} & Western U.S. power grid.~\cite{WS98} & 4941 & 6594 & & 0.612 & 0.804 &  &  & 0.668 & \textbf{0.908} & 1.14 \\
\textit{blogs} & Weblogs on politics.~\cite{AG05} & 1490 & 16718 & & \textbf{0.426} & \textbf{0.426} &  &  & \textbf{0.426} & \textbf{0.426} & 1.00 \\
\textit{pgp} & \textit{PGP} web of trust.~\cite{BPDA04} & 10680 & 24340 & 0.849 & 0.754 & 0.844 & 0.726 & \textbf{\textit{0.884}} & 0.782 & \textbf{0.869} & 1.08 \\
\textit{asi} & Autonomous syst. of Internet.~\cite{New06b} & 22963 & 48436 & & 0.511 & 0.591 & & & 0.528 & \textbf{0.600}\footnotemark[2] & 1.02 & 0 s\\
\textit{codmat$^3$} & \textit{Cond. Matt.} archive 2003.\footnotemark[1]~\cite{New01} & 27519 & 116181 & 0.661 & 0.616 & 0.683 & 0.582 & \textbf{\textit{0.755}} & 0.634 & \textbf{0.735} & 1.00 & 1.5 s\\ 
\textit{codmat$^5$} & \textit{Cond. Matt.} archive 2005.\footnotemark[1]~\cite{New01} & 36458 & 171736 & & 0.586 & 0.643 &  &  & 0.608 & \textbf{0.683} & 1.00 \\
\textit{kdd$^3$} & \textit{KDD-Cup} 2003 dataset.~\cite{Aut03} & 27770 & 352285  & & 0.624 & \textbf{0.630} &  &  & 0.619 & 0.617 & 1.00 & 3 s\\
\textit{nec} & \textit{nec} web overlay map.~\cite{HJJMMMM03} & 75885 & 357317  & & 0.693 & 0.738 &  &  & 0.703 & \textbf{0.767} & 1.03 \\
\textit{epinions} & \textit{Epinions} web of trust.~\cite{RAD03} & 75879 & 508837  & & 0.382 & 0.362 &  &  & 0.399 & \textbf{0.402} & 1.00 & 4.5 s\\ 
\textit{amazon$^3$} & \textit{Amazon} co-purchasing 2003.~\cite{LAH07} & 262111 & 1.2M & & 0.682 & 0.749 &  & & 0.701 & \textbf{0.857} & 1.01 & 20 s\\ 
\textit{ndedu} & Webpages in \textit{nd.edu} domain.~\cite{AJB99} & 325729 & 1.5M & & 0.840 & 0.890 &  & & 0.863 & \textbf{0.903} & 1.14 \\
\textit{google} & Web graph of \textit{Google}.~\cite{LLDM09} & 875713 & 4.3M & & 0.805 & 0.923 &  &  & 0.822 & \textbf{0.968} & 1.01 & 2.5 m\\ 
\textit{nber} & \textit{NBER} patents citations.~\cite{HJT01} & 3.8M & 16.5M & & 0.573 & 0.624 &  &  & 0.583 & \textbf{0.759} & 1.20 \\
\textit{live} & \textit{Live Journal} friendships.~\cite{LLDM09} & 4.8M & 69.0M & & 0.538 & 0.539 &  &  & 0.557 & \textbf{0.693} & 1.00 & 44 m\\ 
\end{tabular}
\end{ruledtabular}
\footnotetext[1]{Reduced to the largest component of the original network.}
\footnotetext[2]{Obtained with slightly modified version of DPA (see caption).}
\footnotetext[3]{Average number of core extractions and computational times for DPA.}
\caption{\label{tbl_eval_rws} Peak (maximal) modularities $Q$ for various label propagation algorithms and a greedy optimization of modularity. The modularity for DPA for \textit{elegans} was obtained with $\delta_{max}=1$ and for \textit{asi} with $\delta_{max}=0$ (\appref{app_algs}); else the values are $0.420$ and $0.588$ respectively. Opaque solid values correspond to the approaches that have significant time complexity compared to DPA.}
\end{table*}

\section{\label{sec_eval}Evaluation and discussion}
The section presents results of the empirical evaluation of the proposed algorithms.

Algorithms were first compared on two classes of benchmark networks with planted partition, namely, \citet{GN02}  and \LFR networks. For the latter, we also vary the size of the networks ($1000$ and $5000$ nodes) and the size of the communities (from $10$ to $50$ and from $20$ to $100$ nodes). Results are assessed in terms of \textit{normalized mutual information} (NMI)~\cite{DDDA05} and are shown in \figref{fig_eval_benchs}.

Analysis clearly shows the difference between defensive and offensive propagation, especially on larger networks (\figref{fig_eval_benchs}~(d,e)). The offensive propagation (ODALPA) performs slightly better than the basic LPA, and can still relatively accurately detect communities, when LPA already performs rather poorly (\figref{fig_eval_benchs}~(d)). On the other hand, the defensive propagation (DDALPA) does not detect communities as accurately as the other two approaches (\figref{fig_eval_benchs}~(d,e)), however, the algorithm still reveals the communities even when they are only weakly defined (and the other two approaches clearly fail). In other words, the defensive algorithm has high \textit{recall}, whereas the offensive approach achieves high \textit{precision}.

Furthermore, BDPA (and DPA) outperforms all three aforementioned algorithms. Note that the performance does not simply equal to the \textit{upper-hull} of those for DDALPA and ODALPA. The analysis also shows that core extraction (i.e. DPA) does not improve the results on networks with thousands of nodes or less; the slight improvement on \GN results only from hierarchical investigation, and not core extraction. Nevertheless, as shown below, the results can be significantly improved on larger networks.

\citet{LF09} have conducted a thorough empirical analysis of more then $10$ state-of-the-art community detection algorithms. To enable the comparison, the benchmark networks in \figref{fig_eval_benchs} were selected so they exactly coincide with those used in~\cite{LF09}. By comparing the results, we can conduct that DPA does indeed perform at least as good as the best algorithms analyzed in~\cite{LF09}, namely, hierarchical modularity optimization of \citet{BGLL08}, \textit{model selection} approach of \citet{RB08}, spectral algorithm proposed by \citet{DM04} and multiresolution \textit{spin model} of \citet{RN10}. Moreover, on larger networks (\figref{fig_eval_benchs}~(d,e)), DPA obtains even better results than all of the algorithms analyzed in~\cite{LF09} -- for $\mu=0.8$, none of the analyzed algorithms can obtain NMI above $\approx0.35$, when the values for DPA are $0.651$, $0.541$ respectively.


DPA (and BDPA) was further analyzed on $23$ real-world networks (\tblref{tbl_eval_rws}), ranging from networks with tens of nodes to networks with several tens of millions of edges~\footnotemark[4]. To conduct a general analysis, we have considered a wide range of different types of real-world networks, in particular, social, communication, citation, collaboration, web, Internet, biological and other networks (all networks were treated as unweighted and undirected.). Due to a large number of networks considered, detailed description is omitted.

\footnotetext[4]{The analysis on networks with hundreds of millions or even billions of edges was bounded due to limited memory resources.}

DPA algorithm was compared with all other proposed label propagation algorithms (due to our knowledge) and a greedy modularity optimization approach (\tblref{tbl_eval_rws}). The algorithms are as follows: LPA denotes basic label propagation~\cite{RAK07} and LPAD denotes LPA with decreasing hop attenuation and node preference equal to the degree of the node~\cite{LHLC09} (\secref{sec_lpa}). The modularity optimization version of LPA is denoted LPAQ~\cite{BC09} and its refinement with multistep greedy merging LPAM~\cite{LM09b}. Furthermore, GMO denotes greedy modularity optimization proposed by~\citet{CNM04}.

For each algorithm, we report peak (maximal) modularities obtained on the networks analyzed. Modularities for LPA, LPAD, BDPA and DPA were obtained by running the algorithms from $2$ to $100000$ times on each network (depends on the size of the network). On the other hand, peak modularities for LPAQ and LPAM (and also GMO) were reported by~\citet{LM09b}.

The results show that DPA outperforms all other label propagation algorithms, except LPAM on networks of medium size (i.e. \textit{elegans}, \textit{emails}, \textit{pgp} and \textit{codmat$^3$}). However, further analysis reveals that, on these networks, LPAM already has considerable time complexity compared to DPA. It should also be noted that modularities, obtained by LPAM on three of these networks, correspond to the highest modularity values ever reported in the literature. Similarly, peak modularities obtained by DPA (and some others) on smaller networks also equal the highest modularities ever published (due to our knowledge, the modularity for \textit{football} even slightly exceeds the highest value ever reported, i.e. $0.606$, opposed to $0.605$).  In summary, DPA obtains significantly higher values of modularity than other comparable label propagation approaches, especially on larger networks (with millions of nodes and edges).

As already discussed in \secref{sec_dpa_main}, BDPA achieves superior results on smaller networks, better than LPA,  LPAQ and LPAD (and GMO). However, the algorithm is not appropriate for larger networks, where hierarchical core extraction prevails (i.e. DPA).

We have also analyzed the number of core extractions (\secref{sec_dpa_main}), made by DPA on these networks (\tblref{tbl_eval_rws}). Core extraction does not gain on networks with less then thousands of nodes or edges, where the average number is commonly close to $0$. However, when networks become larger, a (single) core extraction produces a significant gain in modularity (on these networks). Interestingly, even on the network with several millions of nodes and several tens of millions of edges (i.e. \textit{live}), the number of extractions is still $1$ (on average). 


Next, we have thoroughly compared the time complexity of a simple LPA and DPA (and also LPAM~\cite{LM09b}). On each iteration of the algorithms, each edge of the network is visited (at most) twice. Thus the time complexity of a single iteration equals $\mathcal{O}(m)$, with $m$ being the number of edges. The complexity for DPA is even lower, after the core has been extracted, however, due to simplicity, we consider each iteration to have complexity $\mathcal{O}(m)$.

\begin{figure}
\includegraphics[width=0.925\columnwidth]{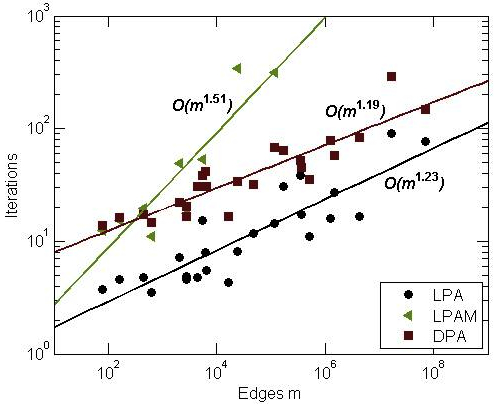}
\caption{\label{fig_eval_time} (Color online) Time complexity of different label propagation algorithms estimated on real-world networks from \tblref{tbl_eval_rws} (results are averages over $100$ iterations). From top to bottom, straight lines correspond to $0.83\mbox{ }m^{0.51}$, $5.15\mbox{ }m^{0.19}$ and $1.03\mbox{ }m^{0.23}$, when the text denotes the overall time complexity of the algorithms  (LPAM, DPA and LPA respectively). On a network with a billion edges, the (projected) number of iterations for DPA, LPA would equal $265$, $113$ respectively.}
\end{figure}

Iterative algorithms (like label propagation) are commonly assessed only on smaller networks, where the number of iterations can be bounded by a small constant. In this context, both LPA and DPA exhibit near linear complexity, $\mathcal{O}(m)$. However, on networks with thousands or millions of nodes and edges, this ``constant'' indeed increases -- even for simple LPA, which is known by its speed, the number of iterations notably increases on larger networks. We have thus analyzed the total number of iterations, made by the algorithms on real-world networks (\tblref{tbl_eval_rws}). The results are shown in \figref{fig_eval_time} (the number of edges $m$ is chosen to represent the size of the network). Note that the number of iterations for DPA corresponds to the sum of the iterations, made by all of the algorithms run within (i.e. DDALPA, ODALPA and BDPA).

\begin{figure*}
\includegraphics[width=2.00\columnwidth]{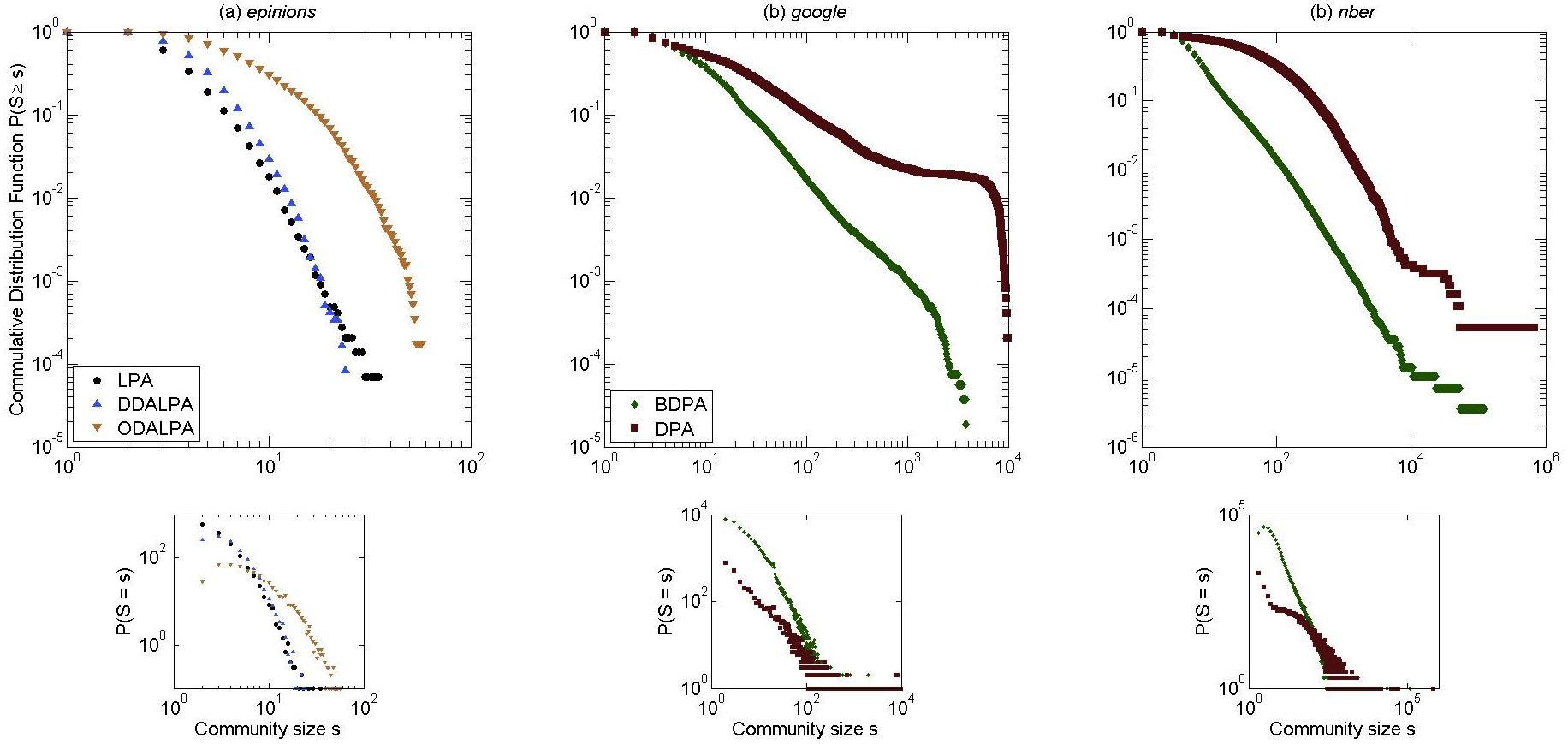}
\caption{\label{fig_eval_sizes} (Color online) (Cumulative) distributions of the community sizes for three real-world networks from \tblref{tbl_eval_rws} (for the \textit{epinions} network, the results were averaged over $10$ runs). Note some particularly large communities revealed by DPA in the case of \textit{google} and \textit{nber} networks (with round $10^4$ and $10^6$ nodes respectively). Interestingly, these coincide with low \textit{conductance}~\cite{Bol98} communities reported in~\cite{LLDM09}.}
\end{figure*}

As already discussed earlier, DPA (and LPA) scale much better than LPAM -- the average number of iterations on the network with tens of millions of edges is $147$, $78$ for DPA, LPA respectively, when LPAM already exceeds $300$ iterations on networks with tens of thousands of edges. Furthermore, results also show that DPA scales even better than simple LPA (i.e. $\mathcal{O}(m^{1.19})$, opposed to $\mathcal{O}(m^{1.23})$), however, it is outperformed by LPA due to a larger constant. Nevertheless, the analysis shows promising results for future analyses of large complex networks.

In the context of analyzing large networks, it should be mentioned that by far the fastest convergence is obtained by using the defensive propagation algorithm DDALPA (\secref{sec_dpa_dalpa}). On the largest of the networks  (i.e. \textit{live}), the algorithm converges in only $25$ iterations (three times faster than LPA), still, the modularity of the revealed community structure is only $0.470$.


Last, we have also studied the stability of DPA (and BDPA), and compare it with simple LPA. The latter is known to find a large number of distinct community structures in each network~\cite{RAK07,TK08,LM09b}, when \citet{TK08} have also argued that these are relatively different between themselves. Indeed, on \textit{zachary} network LPA revealed $628$ different community structures (in $10000$ iterations), when this number equals $159$, $124$ for BDPA, DPA respectively. However, as the number of distinct communities can be misleading, we have rather directly compared the identified community structures.

In~\tblref{tbl_eval_stab} we show mean pairwise NMI of (distinct) community structures that were identified by the algorithms on selected set of real-world networks. DPA (and BDPA) shows to be more stable than LPA, moreover, the identified community structures are relatively similar for all of the algorithms considered (in most networks analyzed). Interestingly, the results also seem to correlate with revealed modularities in~\tblref{tbl_eval_rws} -- clearer the community structure of the network, more stable the algorithms appear. Nevertheless, as indicated by various authors before~\cite{RAK07,TK08}, the number of different community structures can be very high, specially in larger networks (e.g., $1116$, $1330$ for DPA applied to \textit{football}, \textit{jazz} network respectively).

\begin{table}
\begin{ruledtabular}
\begin{tabular}{cccccc}
Network & Nodes & Edges & LPA & BDPA & DPA \\\hline
\textit{karate} & 34 & 78 & 0.574 & 0.578 & \textbf{0.660} \\
\textit{dolphins} & 62 & 159 & 0.714 & 0.762 & \textbf{0.774} \\
\textit{books} & 105 & 441 & 0.737 & 0.803 & \textbf{0.805} \\
\textit{football} & 115 & 616 & 0.878 & 0.896 & \textbf{0.897} \\
\textit{elegans} & 453 & 2025 & 0.610 & 0.615 & \textbf{0.618} \\
\textit{jazz} & 198 & 2742 & 0.602 & 0.748 & \textbf{0.808} \\
\end{tabular}
\end{ruledtabular}
\caption{\label{tbl_eval_stab} Mean pairwise NMI of distinct community structures identified by different label propagation algorithms in $10000$ iterations (on selected set of networks from \tblref{tbl_eval_rws}).}
\end{table}


(Cumulative) distributions of sizes of communities, revealed with the proposed algorithms on three real-world networks, are shown in \figref{fig_eval_sizes}.


\section{\label{sec_conc}Conclusion}
The article proposes an advanced label propagation community detection algorithm that combines two unique strategies of community formation. The algorithm analyzes the network in a hierarchical manner that recursively extracts the core of the network and identifies whisker communities. Algorithm employs only local measures for community detection, and does not require the number of communities to be specified beforehand. The proposition was rigorously analyzed on benchmark networks with planted partition and on a wide range of real-world networks, with up to several millions of nodes and tens of millions of edges. The performance of the algorithm is comparable to the current state-of-the-art community detection algorithms, moreover, the algorithm exhibits almost linear time complexity (in the number of edges of the network) and scales even better than the basic label propagation algorithm. The proposal thus gives prominent grounds for future analysis of large complex networks.

The work also provides further understanding on dynamics of label propagation, in particular, how different propagation strategies can alter the dynamics of the process and reveal community structures, with unique properties. 


\appendix

\begin{acknowledgments}
The authors wish to thank (anonymous) reviewers for comments and criticisms that helped on improving the article.
The work has been supported by the Slovene Research Agency \textit{ARRS} within the research program P2-0359.
\end{acknowledgments}


\section{\label{app_cps}Core-periphery structure}
\citet{LLDM09} have conducted an extensive analysis of large social and information, and some other, networks. They have observed that these networks can be clearly divided into the central \textit{core} and remaining \textit{periphery} (i.e. \textit{core-periphery structure}). Periphery is constituted of many small, well defined communities (in terms of \textit{conductance}~\cite{Bol98}) that are only weakly connected to the rest of the network. When they are connected by a single edge, they are called \textit{whiskers} (or \textit{1-whiskers}). On the other hand, the core of the network consists of larger communities that are well connected between, and thus only loosely defined in the sense of communities. Their analysis have thus revealed that the best communities (due to conductance) reside in the periphery of these networks (i.e. whiskers), and have a characteristic size of around $100$ nodes. For further discussion see~\cite{LLDM09,LLM10}.


\begin{figure}
\algsetup{indent=1em}
\begin{algorithmic}[1]
\REQUIRE Graph $G(N,E)$ with weights $W$
\ENSURE Communities $C$ (i.e. node labels)

\STATE $\delta\gets 0$
\FOR{$n\in N$} 
	\STATE $c_n\gets l_n$ \COMMENT{Unique label.}
	\STATE $d_n\gets 0$
	\STATE $p_n\gets 1/|N|$
\ENDFOR

\STATE \textit{shuffle}$(N)$

\WHILE{\textit{not converged}}
	\FOR{$n\in N$} 
		\STATE $c_n\gets \argmax_l\sum_{i\in\mathcal{N}^l(n)}p_i(1-\delta d_i)w_{ni}$
		\IF{$c_n$ \textit{has changed}}
			\STATE $d_n\gets (\min_{i\in\mathcal{N}^{c_n}(n)}d_i)+1$
			\STATE $p_n\gets \sum_{i\in\mathcal{N}^{c_n}(n)}p_i/k^{c_n}_i$
		\ENDIF
	\ENDFOR
	\STATE $\delta\gets$ \textit{proportion of labels changed}
	\IF{$\delta\geq\delta_{max}$} 
		\STATE \COMMENT{$\delta_{max}$ is fixed to $\frac{1}{2}$.}
		\STATE $\delta\gets 0$
	\ENDIF
\ENDWHILE

\RETURN $C$
\end{algorithmic}
\caption{\label{alg_dalpa}Defensive label propagation algorithm with (dynamic) hop attenuation (DDALPA). In the offensive version (ODALPA), the node preference $p_i$ is replaced by $1-p_i$ (line 10) and the degree $k^{c_n}_i$ is replaced by $k_i$ (line 13).}
\end{figure}

\begin{figure}
\algsetup{indent=1em}
\begin{algorithmic}[0]
\REQUIRE Graph $G(N,E)$ with weights $W$
\ENSURE Communities $C$ (i.e. node labels)

\STATE $C\gets$ \textit{DDALPA}$(G,W)$

\FOR{$c\in C$} 
	\STATE \COMMENT{Retain community cores.}
	\STATE $m_c\gets$\textit{median}$(\{p_n|\mbox{ }n\in N\wedge c_n=c\})$
	\FOR{$n\in N\wedge c_n=c\wedge p_n\leq m_c$} 
		\STATE $c_n\gets l_n$ \COMMENT{Unique label.}
		\STATE $d_n\gets 0$
		\STATE $p_n\gets 0$ \COMMENT{Maximal preference.}
	\ENDFOR
\ENDFOR

\STATE $C\gets$ \textit{ODALPA}$(G,W)$

\RETURN $C$ \COMMENT{Returns best communities.}
\end{algorithmic}
\caption{\label{alg_dpa}Basic diffusion and propagation algorithm (BDPA).}
\end{figure}

\begin{figure}
\algsetup{indent=1em}
\begin{algorithmic}[0]
\REQUIRE Graph $G(N,E)$ with weights $W$
\ENSURE Communities $C$ (i.e. node labels)

\STATE $C\gets$ \textit{DDALPA}$(G, W)$

\STATE $C_C\gets$ \textit{ODALPA}$(G_C, W_C)$

\IF{$C_C$ \textit{contains one community}}
	\STATE $C\gets$ \textit{BDPA}$(G,W)$
\ELSE
	\STATE \COMMENT{Recursion on core $c$ in $C_C$.}
	\STATE $C\gets (C_C-\{c\})$ $\cup$ \textit{DPA}$(G_C(c),W_C(c))$ 
\ENDIF

\RETURN $C$ \COMMENT{Returns best communities.}
\end{algorithmic}
\caption{\label{alg_dpas}Diffusion and propagation algorithm (DPA).}
\end{figure}

\section{\label{app_algs}Algorithms}
In this section we give the pseudo-code of all the algorithms, proposed in the article (\figref{alg_dalpa},  \figref{alg_dpa} and \figref{alg_dpas}), and discuss some of the implementation issues.

Due to the nature of label propagation, it may be that, when the algorithm converges, two (disconnected) communities share the same label. This happens when node propagates its label in two direction, but is itself relabeled in the later stages of the algorithm. Nevertheless, disconnected communities can be detected at the end using a simple \textit{breath-first search}.

Each run of BDPA or DPA (\figref{alg_dpa}, \figref{alg_dpas}) unfolds several sets of communities and the best are returned at the end (due to some measure of goodness of communities). For the analysis in \secref{sec_eval}, algorithms reported community structure that obtained highest modularity (computed on the original network). Thus, the results might be attributed to modularity's \textit{resolution limit} problem~\cite{FB07}, or other limitations~\cite{GMC10}, still, this is not a direct artefact of the algorithms.

Additional note should be made for the offensive propagation algorithm ODALPA (\figref{alg_dalpa}). When used on networks with several thousands of nodes or less, diffusion values $p_n$ should only be updated (line 13) after the first iteration, otherwise the algorithm might not converge. The reason is that, during the first iteration, communities are still rather small (due to the size of the network) and thus all of the nodes lay in the border of the communities. Hence, updating diffusion values results in applying propagation preference to all of the nodes.


%


\end{document}